\documentclass{PoS}

\usepackage{graphicx}
\usepackage{amsmath,amssymb,amsfonts}
\usepackage{psfrag}
\usepackage{accents}
\usepackage{bbm}
\usepackage{bm}
\usepackage{nccbbb}
\usepackage{tikz}
\let\normalcolor\relax

\newcommand{\vc}[1]{{\bm{#1}}}

\newcommand{\ev}[1]{\left\langle #1 \right\rangle}
\newcommand{\nN}[1]{\left\langle #1 \right\rangle}

\newcommand{\abs}[1]{\left| #1 \right|}

\newcommand{\Seff}{S_{\text{eff}}}
\newcommand{\Sw}{S_{\text{W}}}

\newcommand{\pDeriv}[2][]{\frac{\partial #1}{\partial #2}}
\newcommand{\coco}{\mathrm{c.c.}}
\newcommand{\bbZ}{\mathbb{Z}}

\newcommand{\crit}{\mathrm{crit}}
\newcommand{\trP}{\mathcal{P}}

\newcommand{\Ns}{N_\mathrm{s}}
\newcommand{\Nt}{N_\mathrm{t}}
\newcommand{\ord}{\mathcal{O}}
\newcommand{\vcx}{\vc{x}}
\newcommand{\vcy}{\vc{y}}
\newcommand{\vcz}{\vc{z}}

\newcommand{\vcL}{\vc{L}}
\newcommand{\Haar}{\text{Haar}}
\newcommand{\rep}{\mathcal{R}}

\DeclareMathOperator{\tr}{tr}
\DeclareMathOperator{\sgn}{sgn}
\DeclareMathOperator{\diag}{diag}

\renewcommand{\Re}{\mathrm{Re\,}}
\renewcommand{\Im}{\mathrm{Im\,}}

\newcommand{\myindent}{\noindent}

\title{{$\bbbz_\mathbf{3}$} Polyakov Loop
Models and Inverse Monte-Carlo Methods}

\ShortTitle{$\mathbb{Z}_3$ Polyakov Loop
Models and Inverse Monte-Carlo Methods}

\author{\speaker{Christian Wozar}, Tobias K\"astner,
Sebastian Uhlmann, Andreas Wipf\\
        Theoretisch-Physikalisches Institut,
Friedrich-Schiller-Universit{\"a}t Jena, Max-Wien-Platz 1, 07743
Jena, Germany\\
        E-mail: \email{christian.wozar@uni-jena.de}}

\author{Thomas Heinzl\\
       School of Mathematics and Statistics, University of
Plymouth, Drake Circus, Plymouth, PL4~8AA, United Kingdom}

\abstract{We study effective Polyakov loop models for $SU(3)$
Yang-Mills theory at finite temperature. A comprehensive mean
field analysis of the phase diagram is carried out and compared
to the results obtained from Monte-Carlo simulations. We find a
rich phase structure including ferromagnetic and
antiferromagnetic phases. Due to the presence of a tricritical
point the mean field approximation agrees very well with the
numerical data. Critical exponents associated with second-order
transitions coincide with those of the $\bbZ_3$ Potts model.
Finally, we employ inverse Monte-Carlo methods to determine the
effective couplings in order to match the effective models to
Yang-Mills theory.}

\FullConference{The XXV International Symposium on Lattice Field Theory\\
		 July 30-4 August 2007\\
		 Regensburg, Germany}

\begin{document}

% ===========================================================
\section{Introduction}
\myindent
The Svetitsky-Yaffe conjecture
\cite{Svetitsky:1982gs,Falcone:2007rv} states that the
Yang-Mills finite temperature transition in dimension $d+1$ is
described by an effective spin model in $d$ dimensions with
short range interactions. Combining this idea with strong
coupling expansions and inverse Monte-Carlo (IMC) methods we
analyse the relationship between $SU(3)$ YM theory in $3+1$
dimensions and effective theories formulated as $\bbZ_3$ spin
models in $3$ dimensions.

% ===========================================
\section{\bm{$SU(3)$} and characters of representations}
\myindent
Our effective operators are \emph{class functions} on $SU(3)$.
With group elements in diagonal form, $g = \diag (e^{i\phi_1},
e^{i\phi_2}, e^{-i(\phi_1+\phi_2)})$, we associate a group
character in the fundamental representation by
\begin{equation} \label{equ:class}
\trP \equiv \tr g \equiv \chi_{10}(g) =e^{i\phi_1}+e^{i\phi_2}+e^{-i(\phi_1+\phi_2)}
\; ,
\end{equation}
with the typical example being the Polyakov loop. The
parameterisation (\ref{equ:class}) implies the \emph{reduced
Haar measure} on the maximal Abelian torus,
\begin{equation} d\mu_\text{red} = J^2 d\phi_1 d\phi_2, \quad
J^2 = 15-6\chi_{11}+3\chi_{30}+3\chi_{03}-\chi_{22} .
\end{equation}
Using Young tableaux one can express all characters $\chi_{pq}$ with Dynkin
labels $[p,q]$ in terms of the fundamental ones, $\trP$ and $\trP^*$.

% ===========================================
\section{Observables}
\myindent
We discuss YM theory on a $\Ns^3\!\times\!\Nt$-lattice. The
Polyakov loop $\trP_\vcx$ is measured in terms of its lattice
average,
\begin{equation}
P \equiv \frac{1}{V}\sum_\vcx \trP_\vcx,\quad V=\Ns^3.
\end{equation}

The observable relevant for the analysis of antiferromagnetic
phases is
\begin{equation}
M \equiv \frac{1}{V}\sum_\vcx \trP_\vcx\sgn (\vcx), \quad\sgn (\vcx) \equiv
(-1)^{^{\textstyle \sum_i x_i}} \;
\end{equation}
and measures the \emph{difference} of the Polyakov loop on odd
and even sublattices.

Since we will have to deal with phases where the traced
Polyakov loop is located halfway between the $SU(3)$ center
elements we project the value of the traced Polyakov loop onto
the nearest $\bbZ_3$-axis and define a \emph{rotated Polyakov loop} by\\
\begin{minipage}{0.49\linewidth} \begin{equation}
 P_r =
 \begin{cases}
  \phantom{-\frac{1}{2}}\Re P \; & :\;
  P\in \mathcal{F} \\
  -\frac{1}{2}\Re P + \frac{\sqrt{3}}{2}\Im P \; & :\;
  P\in \mathcal{F'}  \\
  -\frac{1}{2}\Re P-\frac{\sqrt{3}}{2}\Im P \; & :\;
  P\in \mathcal{F''}
  \end{cases} .
\end{equation}
\end{minipage}
\begin{minipage}{0.5\linewidth}
\center{\vspace*{3pt}
% This file is generated by the MATLAB m-file laprint.m. It can be included
% into LaTeX documents using the packages graphicx, color and psfrag.
% It is accompanied by a postscript file. A sample LaTeX file is:
%    \documentclass{article}\usepackage{graphicx,color,psfrag}
%    \begin{document}\input{rotatedObs}\end{document}
% See http://www.mathworks.de/matlabcentral/fileexchange/loadFile.do?objectId=4638
% for recent versions of laprint.m.
%
% created by:           LaPrint version 3.16 (13.9.2004)
% created on:           08-May-2006 11:05:11
% eps bounding box:     15 cm x 11.25 cm
% comment:              
%
\begin{psfrags}%
\psfragscanon%
%
% text strings:
\psfrag{s03}[l][l]{\color[rgb]{0,0,0}\setlength{\tabcolsep}{0pt}\begin{tabular}{l}\footnotesize$\mathcal{F}$\end{tabular}}%
\psfrag{s04}[l][l]{\color[rgb]{0,0,0}\setlength{\tabcolsep}{0pt}\begin{tabular}{l}\footnotesize$\mathcal{F}$\end{tabular}}%
\psfrag{s05}[l][l]{\color[rgb]{0,0,0}\setlength{\tabcolsep}{0pt}\begin{tabular}{l}\footnotesize$\mathcal{F'}$\end{tabular}}%
\psfrag{s06}[l][l]{\color[rgb]{0,0,0}\setlength{\tabcolsep}{0pt}\begin{tabular}{l}\footnotesize$\mathcal{F''}$\end{tabular}}%
\psfrag{s07}[c][l]{\color[rgb]{0,0,0}\setlength{\tabcolsep}{0pt}\begin{tabular}{l}\footnotesize$\mathcal{F''}$\end{tabular}}%
\psfrag{s08}[Br][Bc]{\color[rgb]{0,0,0}\setlength{\tabcolsep}{0pt}\begin{tabular}{l}\footnotesize$\mathcal{F'}$\end{tabular}}%
\psfrag{s09}[t][t]{\color[rgb]{0,0,0}\setlength{\tabcolsep}{0pt}\begin{tabular}{c}\footnotesize$\Re
P$\end{tabular}}%
\psfrag{s10}[b][b][1][270]{\color[rgb]{0,0,0}\setlength{\tabcolsep}{0pt}\begin{tabular}{c}\footnotesize$\Im
P$\end{tabular}}% % % xticklabels:
\psfrag{x01}[t][t]{\scriptsize$-2$}%
\psfrag{x02}[t][t]{\scriptsize$-1$}%
\psfrag{x03}[t][t]{\scriptsize$0$}%
\psfrag{x04}[t][t]{\scriptsize$1$}%
\psfrag{x05}[t][t]{\scriptsize$2$}%
\psfrag{x06}[t][t]{\scriptsize$3$}%
\psfrag{x07}[t][t]{\scriptsize$4$}%
%
% yticklabels:
\psfrag{v01}[r][r]{\scriptsize\phantom{$-2.5$}}%
\psfrag{v02}[r][r]{\scriptsize$-2$}%
\psfrag{v03}[r][r]{\scriptsize\phantom{$-1.5$}}%
\psfrag{v04}[r][r]{\scriptsize$-1$}%
\psfrag{v05}[r][r]{\scriptsize\phantom{$-0.5$}}%
\psfrag{v06}[r][r]{\scriptsize$0$}%
\psfrag{v07}[r][r]{\scriptsize\phantom{$0.5$}}%
\psfrag{v08}[r][r]{\scriptsize$1$}%
\psfrag{v09}[r][r]{\scriptsize\phantom{$1.5$}}%
\psfrag{v10}[r][r]{\scriptsize$2$}%
\psfrag{v11}[r][r]{\scriptsize\phantom{$2.5$}}%
%
% Figure:
\includegraphics[width=0.8\linewidth]{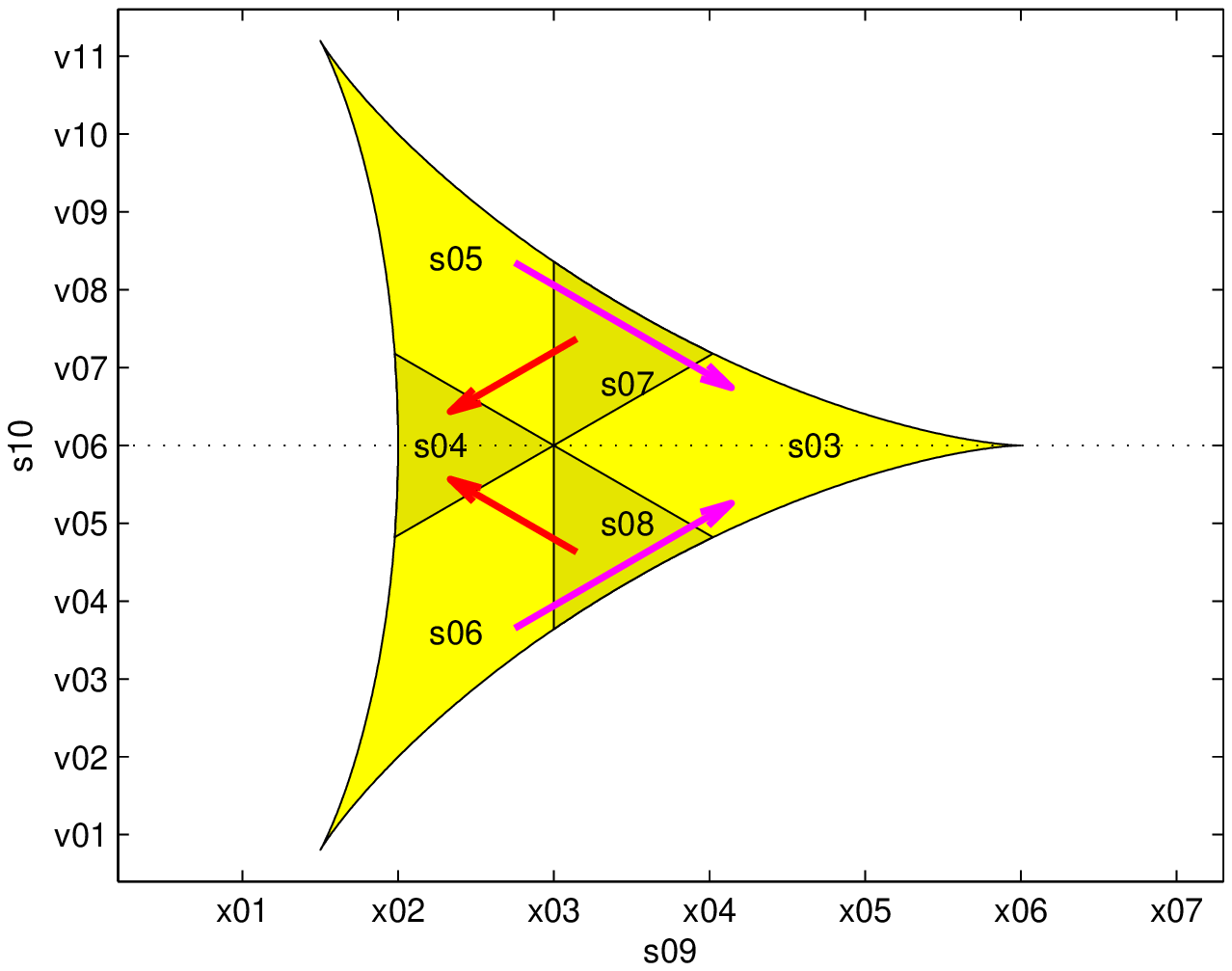}%
\end{psfrags}%
%
% End rotatedObs.tex

}\end{minipage}

% ===========================================
\section{Effective models for Yang-Mills theory}
\myindent
We start with the well-known lattice Wilson action
\begin{equation}
 \Sw = \beta \sum_{\square} \left( 1-  \frac{1}{N_C} \Re \, \tr
 \, U_{\square} \right),\quad \beta = \frac{6}{a^4g^2}
\end{equation}
and perform a \emph{strong coupling expansion} (for small
$\beta$). Since the resulting `operators' (Polyakov loop
monomials) are \emph{dimensionless} there is no natural
ordering scheme. We therefore use a truncation scheme based on:
\begin{itemize}
  \item Ordering by powers of $\beta$ which are closely
      related to the dimension of the corresponding group
      representations.
  \item Ordering by the distance across which the Polyakov
      loops are coupled.
\end{itemize}
In compact form the strong coupling expansion is given by
\begin{equation}
  \Seff = \sum_r \sum_{\rep_1 \ldots  \rep_r} \sum_{\ell_1 \ldots \ell_r}
  c_{\rep_1 \ldots \rep_r}^{\ell_1 \ldots  \ell_r}(\beta)  \prod_{i=1}^r
  S_{\rep_i, \ell_i} = \sum_i \lambda_i S_i
\end{equation}
with the basic building blocks
\begin{equation}
  S_{\rep, \ell} \equiv \chi_\rep (\trP_\vcx) \chi_\rep^* (\trP_{\vcy}) + \coco
  , \quad \ell \equiv \nN{\vcx\vcy} .
\end{equation}
Here $r$ counts the number of link operators contributing at
each order. The coefficients $c_{\rep_1 \ldots \rep_r}^{\ell_1
\ldots \ell_r}$ are the couplings between the operators
$S_{\rep_i,\ell_i}$ sitting at nearest-neighbor (NN) links
$\ell_i \equiv \nN{\vcx_i,\vcy_i}$ in representation $\rep_i$.
The effective action hence describes a \emph{network of link
operators} that are collected into (possibly disconnected)
`polymers' contributing with `weight' $c_{\rep_1 \ldots
\rep_r}^{\ell_1 \ldots \ell_r}$. One expects the `weights' or
couplings to decrease as the dimensions of the involved
representations and inter-link distances increase. In a strong
coupling (small $\beta$) expansion truncated at $\ord(\beta^{k
\Nt})$ one has $r \le k$ and the additional restriction
$|\rep_1| + \cdots + |\rep_r| < k$ with $|\rep| \equiv p+q $
for a given representation $\rep$ with Dynkin labels $[p,q]$.

% ===========================================
\section{A toy model -- mean field vs.\ Monte-Carlo}
\myindent
We consider the $SU(3)$ model \cite{Wozar:2006fi,Wipf:2006wj}
\begin{equation}\label{equ:effModel}
S = \lambda_1 \sum_{\nN{\vcx\vcy}}\bigl(\chi_{10}(\trP_\vcx)\chi_{01}(\trP_\vcy)+\coco\bigr)
 +
\lambda_4 \sum_{\nN{\vcx\vcy}}\bigl(\chi_{10}(\trP_\vcx)\chi_{20}(\trP_\vcy)+ \chi_{20}(\trP_\vcx)\chi_{10}(\trP_\vcy)
+\coco \bigr) .
\end{equation}

\begin{figure}
\center{
% This file is generated by the MATLAB m-file laprint.m. It can be included
% into LaTeX documents using the packages graphicx, color and psfrag.
% It is accompanied by a postscript file. A sample LaTeX file is:
%    \documentclass{article}\usepackage{graphicx,color,psfrag}
%    \begin{document}\input{unnamed}\end{document}
% See http://www.mathworks.de/matlabcentral/fileexchange/loadFile.do?objectId=4638
% for recent versions of laprint.m.
%
% created by:           LaPrint version 3.16 (13.9.2004)
% created on:           24-Jul-2007 15:56:35
% eps bounding box:     15 cm x 11.25 cm
% comment:              
%
\begin{psfrags}%
\psfragscanon%
%
% text strings:
\psfrag{s05}[t][t]{\color[rgb]{0,0,0}\setlength{\tabcolsep}{0pt}\begin{tabular}{c}\footnotesize$\lambda_1$\end{tabular}}%
\psfrag{s06}[r][c][1][270]{\color[rgb]{0,0,0}\setlength{\tabcolsep}{0pt}\begin{tabular}{c}\footnotesize$\lambda_4\;\;$\end{tabular}}%
\psfrag{s09}[][]{\color[rgb]{0,0,0}\setlength{\tabcolsep}{0pt}\begin{tabular}{c} \end{tabular}}%
\psfrag{s10}[][]{\color[rgb]{0,0,0}\setlength{\tabcolsep}{0pt}\begin{tabular}{c} \end{tabular}}%
\psfrag{s11}[b][t][1][270]{\color[rgb]{0,0,0}\setlength{\tabcolsep}{0pt}\begin{tabular}{c}\footnotesize$\;\;\quad
P_\mathrm{r}$\end{tabular}}% %
% xticklabels:
\psfrag{x01}[t][t]{\scriptsize$0.0$}%
\psfrag{x02}[t][t]{\scriptsize$0.5$}%
\psfrag{x03}[t][t]{\scriptsize$1.0$}%
\psfrag{x04}[t][t]{\scriptsize$-0.2$}%
\psfrag{x05}[t][t]{\scriptsize$-0.1$}%
\psfrag{x06}[t][t]{\scriptsize$0.0$}%
\psfrag{x07}[t][t]{\scriptsize$0.1$}%
\psfrag{x08}[t][t]{\scriptsize$0.2$}%
\psfrag{x09}[t][t]{\scriptsize$0.3$}%
%
% yticklabels:
\psfrag{v01}[l][l]{\scriptsize$-1.0$}%
\psfrag{v02}[l][l]{\scriptsize$-0.5$}%
\psfrag{v03}[l][l]{\scriptsize$0.0$}%
\psfrag{v04}[l][l]{\scriptsize$0.5$}%
\psfrag{v05}[l][l]{\scriptsize$1.0$}%
\psfrag{v06}[l][l]{\scriptsize$1.5$}%
\psfrag{v07}[l][l]{\scriptsize$2.0$}%
\psfrag{v08}[l][l]{\scriptsize$2.5$}%
\psfrag{v09}[l][l]{\scriptsize$3.0$}%
\psfrag{v10}[r][r]{\scriptsize$-0.20$}%
\psfrag{v11}[r][r]{\scriptsize$-0.15$}%
\psfrag{v12}[r][r]{\scriptsize$-0.10$}%
\psfrag{v13}[r][r]{\scriptsize$-0.05$}%
\psfrag{v14}[r][r]{\scriptsize$0.00$}%
\psfrag{v15}[r][r]{\scriptsize$0.05$}%
\psfrag{v16}[r][r]{\scriptsize$0.10$}%
\psfrag{v17}[r][r]{\scriptsize$0.15$}%
%
% Figure:
\includegraphics[width=0.45\linewidth]{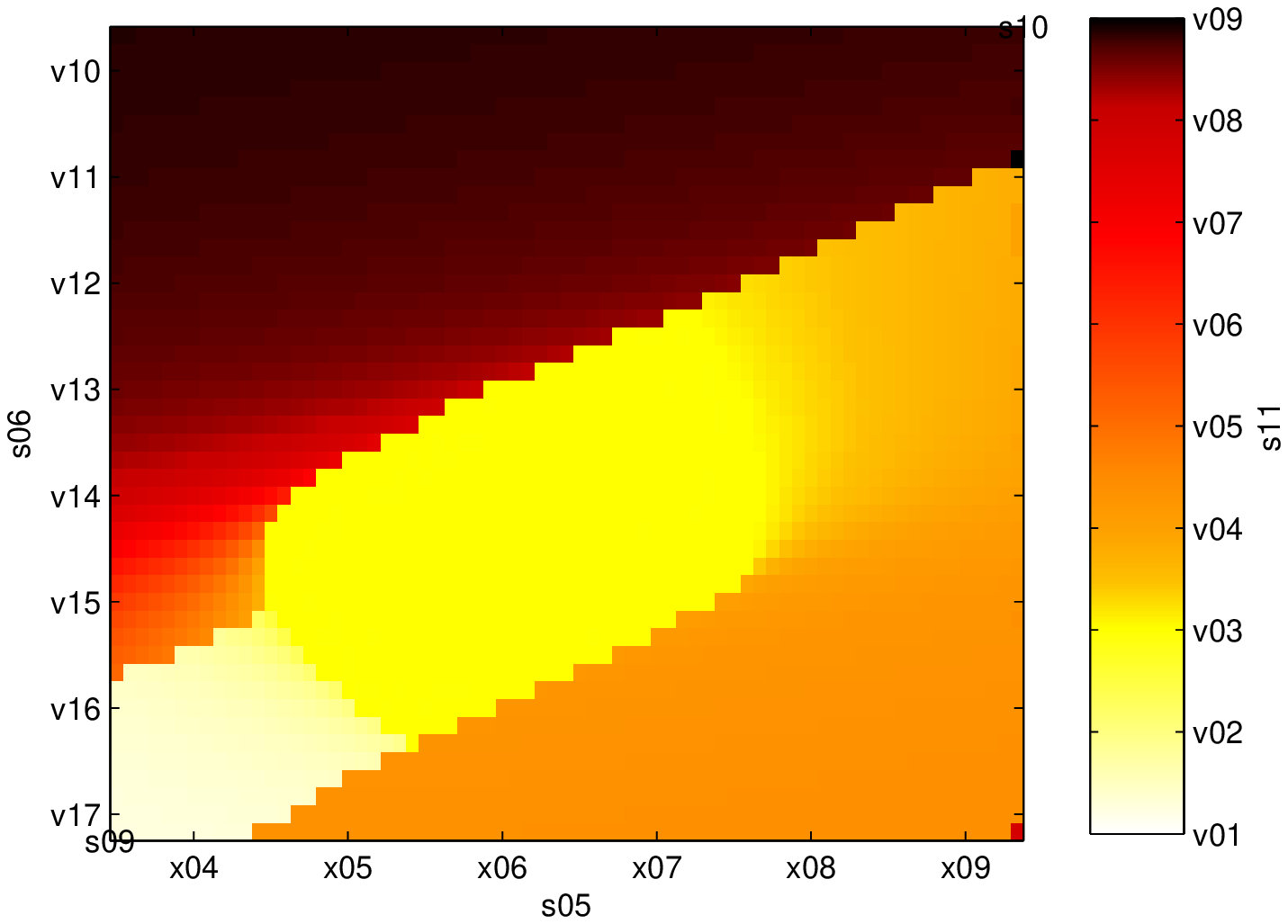}%
\end{psfrags}%
%
% End unnamed.tex
\quad% This file is generated by the MATLAB m-file laprint.m. It can be included
% into LaTeX documents using the packages graphicx, color and psfrag.
% It is accompanied by a postscript file. A sample LaTeX file is:
%    \documentclass{article}\usepackage{graphicx,color,psfrag}
%    \begin{document}\input{unnamed}\end{document}
% See http://www.mathworks.de/matlabcentral/fileexchange/loadFile.do?objectId=4638
% for recent versions of laprint.m.
%
% created by:           LaPrint version 3.16 (13.9.2004)
% created on:           24-Jul-2007 16:05:02
% eps bounding box:     15 cm x 11.25 cm
% comment:              
%
\begin{psfrags}%
\psfragscanon%
%
% text strings:
\psfrag{s05}[t][t]{\color[rgb]{0,0,0}\setlength{\tabcolsep}{0pt}\begin{tabular}{c}\footnotesize$\lambda_1$\end{tabular}}%
\psfrag{s06}[r][c][1][270]{\color[rgb]{0,0,0}\setlength{\tabcolsep}{0pt}\begin{tabular}{c}\footnotesize$\lambda_4\;\;$\end{tabular}}%
\psfrag{s09}[][]{\color[rgb]{0,0,0}\setlength{\tabcolsep}{0pt}\begin{tabular}{c} \end{tabular}}%
\psfrag{s10}[][]{\color[rgb]{0,0,0}\setlength{\tabcolsep}{0pt}\begin{tabular}{c} \end{tabular}}%
\psfrag{s11}[t][t][1][270]{\color[rgb]{0,0,0}\setlength{\tabcolsep}{0pt}\begin{tabular}{c}\footnotesize$\;\;\quad
P_\mathrm{r}$\end{tabular}}% % % xticklabels:
\psfrag{x01}[t][t]{\scriptsize$0.0$}%
\psfrag{x02}[t][t]{\scriptsize$0.5$}%
\psfrag{x03}[t][t]{\scriptsize$1.0$}%
\psfrag{x04}[t][t]{\scriptsize$-0.2$}%
\psfrag{x05}[t][t]{\scriptsize$-0.1$}%
\psfrag{x06}[t][t]{\scriptsize$0.0$}%
\psfrag{x07}[t][t]{\scriptsize$0.1$}%
\psfrag{x08}[t][t]{\scriptsize$0.2$}%
\psfrag{x09}[t][t]{\scriptsize$0.3$}%
%
% yticklabels:
\psfrag{v01}[l][l]{\scriptsize$-1.0$}%
\psfrag{v02}[l][l]{\scriptsize$-0.5$}%
\psfrag{v03}[l][l]{\scriptsize$0.0$}%
\psfrag{v04}[l][l]{\scriptsize$0.5$}%
\psfrag{v05}[l][l]{\scriptsize$1.0$}%
\psfrag{v06}[l][l]{\scriptsize$1.5$}%
\psfrag{v07}[l][l]{\scriptsize$2.0$}%
\psfrag{v08}[l][l]{\scriptsize$2.5$}%
\psfrag{v09}[l][l]{\scriptsize$3.0$}%
\psfrag{v10}[r][r]{\scriptsize$-0.20$}%
\psfrag{v11}[r][r]{\scriptsize$-0.15$}%
\psfrag{v12}[r][r]{\scriptsize$-0.10$}%
\psfrag{v13}[r][r]{\scriptsize$-0.05$}%
\psfrag{v14}[r][r]{\scriptsize$0.00$}%
\psfrag{v15}[r][r]{\scriptsize$0.05$}%
\psfrag{v16}[r][r]{\scriptsize$0.10$}%
\psfrag{v17}[r][r]{\scriptsize$0.15$}%
%
% Figure:
\includegraphics[width=0.46\linewidth]{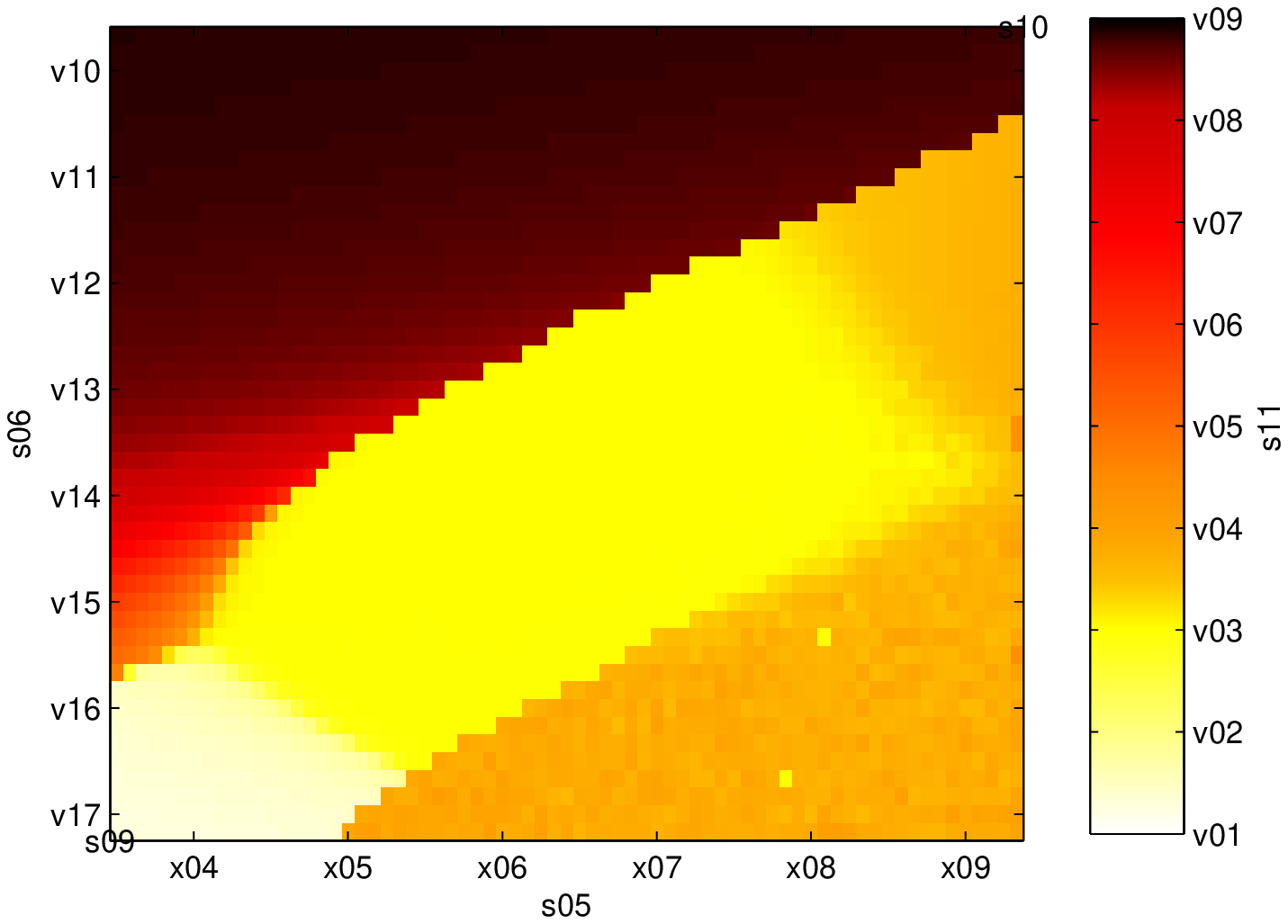}%
\end{psfrags}%
%
% End unnamed.tex

}
\caption{\label{fig:phaseDiags} Phase diagrams obtained by mean field analysis
\textsl{(left)} and Monte-Carlo simulation \textsl{(right)}.}
\end{figure}

A mean field approximation can be applied to approximately
determine the associated phase diagram. We use the following
ansatz for the distribution $p$ of the field $\trP$,
\begin{equation}
p[\mathcal{P}] \to p_{\text{mf}}[\mathcal{P}] \equiv
\prod_{\vcx} p_{\vcx} (\mathcal{P}_{\vcx})\quad
\text{with}\quad   p_{\vcx}(\mathcal{P}_{\vcx}) =
  \begin{cases}
  p_{\mathrm{e}}(\mathcal{P}_{\vcx}) \; &:\; \sgn (\vcx) =  1 \\
  p_{\mathrm{o}}(\mathcal{P}_{\vcx}) \; &:\; \sgn (\vcx) = -1
  \end{cases}.
\end{equation}
The resulting phase diagram is displayed in
Fig.~\ref{fig:phaseDiags} (left panel).

A straightforward Monte-Carlo simulation on an $8^3$-lattice
with a Metropolis algorithm using our \texttt{jenLaTT} package
leads to a phase diagram (Fig.~\ref{fig:phaseDiags}, right
panel) similar to the one obtained by the mean field analysis.
This agreement is due to the presence of a \emph{tri-critical
point} implying an upper critical dimension of three. In
summary, the full phase structure consists of a
\emph{symmetric} phase (in the center of each panel of
Fig.~\ref{fig:phaseDiags}), a \emph{ferromagnetic} phase (upper
left), an \emph{anti-center}  phase (lower left) and an
\emph{antiferromagnetic} (lower right) phase. The anti-center
phase is related to the `skewed' phase of \cite{Myers:2007vc}.

% ===========================================
\section{Algorithms for first and second order transitions}
\myindent
The simulations for the \emph{microscopic YM theory} were done using standard heat
bath algorithms. For the \emph{effective models} we employed standard
Metropolis updates to find the phase diagram. In the vicinity
of phase transitions we made use of the following specially
designed update scheme.

For \emph{first order} transitions we used a multicanonical
algorithm \cite{Berg:1998nj} improving the transition rate
near critical points. For larger lattices the distribution
$\rho$ of the order parameter (denoted $\ell$) was predicted
using the scaling relation
\begin{equation}
 \log \rho (\ell,V) \approx A(\ell) + C(\ell) \,  V \; .
\end{equation}

For \emph{second order} phase transitions algorithms of Wolff
or Swendsen-Wang type lead to strong suppression of the
dynamical critical exponent. These algorithms are useful for
systems with involutory global symmetries, where the local
application of such symmetries leads to ergodic behavior of the
system. For our system there is \emph{no ergodic symmetry}. So
we had to modify the well-known Wolff cluster algorithm
\cite{Wolff:1988uh} as follows:
\begin{enumerate}
\item Choose a random number $N_M$ between $0$ and $V = N^3$.
\item Do $N_M$ standard Metropolis sweeps at randomly drawn lattice points.
\item For a suitable fixed number $N_{\mathrm{cl}}$ repeat the
steps for building a cluster by using the complex conjugation symmetry and its
$\bbZ_3$-symmetric equivalents.
\item Do $V - N_M$ additional Metropolis sweeps, again at randomly chosen
lattice sites.
\end{enumerate}

% ===========================================
\section{Critical exponents for the antiferromagnetic \bm{$SU(3)$} model}
\myindent
For the model \eqref{equ:effModel} with $\lambda_4=0$ we
observe a second order transition between symmetric and
antiferromagnetic phase. Critical exponents $\nu$ and $\gamma$
may be introduced in terms of the relations
\begin{equation}
  \chi(\lambda_{1,\crit}) \propto N^{\gamma/\nu}, \quad
  \left.\frac{\partial U(N,\lambda_{1})}{\partial
  \lambda_{1}}\right|_{\lambda_{1}=\lambda_{1,\crit}} \propto
  N^{1/\nu}
\quad\text{with}\quad
  U = 1 - \frac{\ev{M^4}}{3\ev{M^2}^2} , \quad
  \chi = N^3\ev{M^2}.
\end{equation}
A Monte-Carlo simulation with our modified Wolff cluster
algorithm leads to the following critical exponents in
comparison to the $\bbZ_3$ Potts values:
\begin{center}
 \begin{tabular}{l@{$\qquad$}c@{$\qquad$}c} \hline \hline
  exponent & $\bbZ_3$ Potts \cite{Gottlob:1994ds} & minimal Polyakov\\ \hline
  $\nu$ & $0.664(4)$ & $0.68(2)$ \\
  $\gamma/\nu$ & $1.973(9)$ & $1.96(2)$ \\
  \hline \hline
 \end{tabular}
\end{center}
As the exponents coincide (up to statistical errors) the
$SU(3)$ model is indeed in the same universality class as the
$\bbZ_3$ Potts model (the $XY$ universality class).

% ===========================================
\section{Inverse Monte-Carlo -- the basics}
\myindent
The inverse Monte-Carlo (IMC) method as designed in
\cite{Falcioni:1985bh} allows to determine (effective) actions
from given configurations. In our case, these are Polyakov
loops obtained from gauge configurations generated with the
Wilson action. Via IMC we want to determine the couplings of
truncated effective actions which (ideally) would give rise to
the same distribution of Polyakov loop configurations.

The IMC procedure is based on an \emph{ansatz} for the
effective action of the type $\Seff = \sum_i \lambda_i S_i$.
Translational invariance of the reduced Haar measure leads to
Schwinger-Dyson equations (SDE), see below. They constitute an
\emph{overdetermined} linear system for the effective couplings
$\lambda_i$ which may be solved by least-square methods. As a
further technical input we require a suitable normalization
procedure to make sure that individual equations are
appropriately weighted \cite{Wozar:2007tz}.

% ===========================================
\section{Geometric SDE from invariant group integrals}
\myindent
Translational invariance of the Haar measure implies that
\begin{equation} \label{equ:GSDE}
\int d\mu_\Haar(g) (L_af)(g) = 0\quad \text{for } f\in L_2(G)
\end{equation}
with $L_a$ being the left derivative on the group. Choosing
$f=FL^a \chi_p$ with a class function $F$ and a fundamental
character $\chi_p$ \cite{Uhlmann:2006ze} one obtains
\begin{equation}
L_a(FL^a\chi_p) = F\vcL^2 \chi_p + (L_a F)(L^a \chi_p) ,
\end{equation}
and (\ref{equ:GSDE}) reduces to
\begin{equation}
0=\int d\mu_\Haar \left(F \vcL^2 \chi_p +\sum_q (L_a \chi_p)\frac{\partial F}{\partial
\chi_q } (L^a \chi_q)\right). \end{equation}
Making use of $\vcL^2 \chi_\mu = -c_\mu \chi_\mu$ and of
\begin{equation}
(L_a \chi_\mu)(L^a \chi_\nu) = \frac{1}{2}(c_\mu+c_\nu)\chi_\mu
\chi_\nu -\frac{1}{2}\sum_\lambda C_{\mu\nu}^\lambda	
c_\lambda \chi_\lambda \end{equation} with Casimir values
$c_\mu$ and Clebsch-Gordan coefficients $C_{\mu\nu}^\lambda$
the equation can be specialized to the case of $SU(3)$. For a
suitable function $F$, the SDE finally become
\begin{equation}
\begin{split}
0 &= \ev{-\frac{16}{3} \trP_\vcz S_{i,\trP_\vcx} + (4\trP_\vcz^*-\frac{4}{3}\trP_\vcz^2)
S_{i,\trP_\vcx,\trP_\vcz} +
(6-\frac{2}{3}\abs{\trP_\vcz}^2) S_{i,\trP_\vcx,\trP^*_\vcz}}\\
&\quad - \sum_j \lambda_j\ev{(4\trP_\vcz^*-\frac{4}{3}\trP_\vcz^2) S_{i,\trP_\vcx}
S_{j,\trP_\vcz}
+(6-\frac{2}{3}\abs{\trP_\vcz}^2) S_{i,\trP_\vcx} S_{j,\trP^*_\vcz} } .
\end{split}
\end{equation}

% ===========================================
\section{Algebraic SDE}
\myindent
For $SU(3)$ (generalizations for $SU(N)$ are possible
\cite{Uhlmann:2006ze}) we have the identity
\begin{equation} \int_\Omega d\trP \, d{\trP^*}\, \partial_\trP
f = 0
\end{equation}
which holds for any function $f$ vanishing on $\partial\Omega$.
Choosing
\begin{equation}
f(\trP,{\trP^*}) = J^3 g(\trP,{\trP^*}),\quad
g_\vcx = \pDeriv[h]{\trP^*_\vcx}\exp(-S),\quad
h = S_i
\end{equation}
we obtain the `algebraic SDE'
\begin{equation}
0 = \ev{\frac{3}{2}\pDeriv[J_\vcz^2]{\trP_\vcz} {S_i}_{,\trP^*_\vcx}
+ J_\vcz^2 S_{i,\trP^*_\vcx,\trP_\vcz}}-\sum_j \lambda_j
\ev{J_\vcz^2 S_{i,\trP^*_\vcx} S_{j,\trP_\vcz} }.
\end{equation}

% ===========================================
\section{IMC results}
\begin{floatingfigure}[r]
% This file is generated by the MATLAB m-file laprint.m. It can be included
% into LaTeX documents using the packages graphicx, color and psfrag.
% It is accompanied by a postscript file. A sample LaTeX file is:
%    \documentclass{article}\usepackage{graphicx,color,psfrag}
%    \begin{document}\input{polFit16}\end{document}
% See http://www.mathworks.de/matlabcentral/fileexchange/loadFile.do?objectId=4638
% for recent versions of laprint.m.
%
% created by:           LaPrint version 3.16 (13.9.2004)
% created on:           13-Apr-2007 16:08:48
% eps bounding box:     15 cm x 11.25 cm
% comment:              
%
\begin{psfrags}%
\psfragscanon%
%
% text strings:
\psfrag{s05}[t][t]{\color[rgb]{0,0,0}\setlength{\tabcolsep}{0pt}\begin{tabular}{c}\small$\beta$\end{tabular}}%
\psfrag{s06}[c][b][1][270]{\color[rgb]{0,0,0}\setlength{\tabcolsep}{0pt}\begin{tabular}{c}\small$\ev{\abs{P}}$\end{tabular}}%
\psfrag{s10}[][]{\color[rgb]{0,0,0}\setlength{\tabcolsep}{0pt}\begin{tabular}{c} \end{tabular}}%
\psfrag{s11}[][]{\color[rgb]{0,0,0}\setlength{\tabcolsep}{0pt}\begin{tabular}{c} \end{tabular}}%
\psfrag{s12}[l][l]{\color[rgb]{0,0,0}geometric}%
\psfrag{s13}[l][l]{\color[rgb]{0,0,0}\scriptsize YM}%
\psfrag{s14}[l][l]{\color[rgb]{0,0,0}\scriptsize algebraic}%
\psfrag{s15}[l][l]{\color[rgb]{0,0,0}\scriptsize geometric}%
%
% xticklabels:
\psfrag{x01}[t][t]{$0$}%
\psfrag{x02}[t][t]{$0.1$}%
\psfrag{x03}[t][t]{$0.2$}%
\psfrag{x04}[t][t]{$0.3$}%
\psfrag{x05}[t][t]{$0.4$}%
\psfrag{x06}[t][t]{$0.5$}%
\psfrag{x07}[t][t]{$0.6$}%
\psfrag{x08}[t][t]{$0.7$}%
\psfrag{x09}[t][t]{$0.8$}%
\psfrag{x10}[t][t]{$0.9$}%
\psfrag{x11}[t][t]{$1$}%
\psfrag{x12}[t][t]{\footnotesize$4.5$}%
\psfrag{x13}[t][t]{\footnotesize$5.0$}%
\psfrag{x14}[t][t]{\footnotesize$5.5$}%
\psfrag{x15}[t][t]{\footnotesize$6.0$}%
\psfrag{x16}[t][t]{\footnotesize$6.5$}%
\psfrag{x17}[t][t]{\footnotesize$7.0$}%
\psfrag{x18}[t][t]{\footnotesize$7.5$}%
\psfrag{x19}[t][t]{\footnotesize$8.0$}%
%
% yticklabels:
\psfrag{v01}[r][r]{$0$}%
\psfrag{v02}[r][r]{$0.1$}%
\psfrag{v03}[r][r]{$0.2$}%
\psfrag{v04}[r][r]{$0.3$}%
\psfrag{v05}[r][r]{$0.4$}%
\psfrag{v06}[r][r]{$0.5$}%
\psfrag{v07}[r][r]{$0.6$}%
\psfrag{v08}[r][r]{$0.7$}%
\psfrag{v09}[r][r]{$0.8$}%
\psfrag{v10}[r][r]{$0.9$}%
\psfrag{v11}[r][r]{$1$}%
\psfrag{v12}[r][r]{\footnotesize$0.0$}%
\psfrag{v13}[r][r]{\footnotesize$0.2$}%
\psfrag{v14}[r][r]{\footnotesize$0.4$}%
\psfrag{v15}[r][r]{\footnotesize$0.6$}%
\psfrag{v16}[r][r]{\footnotesize$0.8$}%
\psfrag{v17}[r][r]{\footnotesize$1.0$}%
\psfrag{v18}[r][r]{\footnotesize$1.2$}%
\psfrag{v19}[r][r]{\footnotesize$1.4$}%
%
% Figure:
\includegraphics[width=0.5\linewidth]{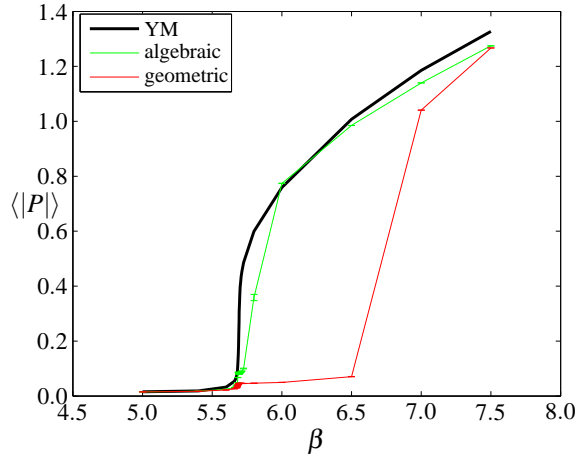}%
\end{psfrags}%
%
% End polFit16.tex

\caption{\label{fig:polFit16}Results obtained from algebraic and geometric SD
equations compared to YM results.}
\end{floatingfigure}
\myindent
We have simulated the underlying YM theory with heat bath
methods on different lattice sizes near the critical coupling.
The corresponding couplings for the effective models were then
determined via IMC \cite{Wozar:2007tz}. The IMC codes were
checked by simulating effective theories with given input
couplings\linebreak which were consistently reproduced by the IMC
procedure. In these tests the algebraic and geometric Ward
identities led to comparable results, limited only by the
statistical accuracy.

We compared the Polyakov loop arising from simulations of full
YM theory and from effective actions based on both geometric
and algebraic SDE (on a $16^3\!\times\!4$-lattice). We found
that the algebraic identities outperformed the geometric ones
in reproducing the YM critical behavior, in particular the
critical coupling (Fig.~\ref{fig:polFit16}).

\begin{figure}
\centering
% This file is generated by the MATLAB m-file laprint.m. It can be included
% into LaTeX documents using the packages graphicx, color and psfrag.
% It is accompanied by a postscript file. A sample LaTeX file is:
%    \documentclass{article}\usepackage{graphicx,color,psfrag}
%    \begin{document}\input{couplingsNN}\end{document}
% See http://www.mathworks.de/matlabcentral/fileexchange/loadFile.do?objectId=4638
% for recent versions of laprint.m.
%
% created by:           LaPrint version 3.16 (13.9.2004)
% created on:           18-Apr-2007 12:27:41
% eps bounding box:     15 cm x 11.25 cm
% comment:              
%
\begin{psfrags}%
\psfragscanon%
%
% text strings:
\psfrag{s05}[t][t]{\color[rgb]{0,0,0}\setlength{\tabcolsep}{0pt}\begin{tabular}{c}\small$\beta$\end{tabular}}%
\psfrag{s10}[][]{\color[rgb]{0,0,0}\setlength{\tabcolsep}{0pt}\begin{tabular}{c} \end{tabular}}%
\psfrag{s11}[][]{\color[rgb]{0,0,0}\setlength{\tabcolsep}{0pt}\begin{tabular}{c} \end{tabular}}%
\psfrag{s12}[l][l]{\color[rgb]{0,0,0}\small$\lambda_{11}$}%
\psfrag{s13}[l][l]{\color[rgb]{0,0,0}\small$\lambda_1$}%
\psfrag{s14}[l][l]{\color[rgb]{0,0,0}\small$\lambda_2$}%
\psfrag{s15}[l][l]{\color[rgb]{0,0,0}\small$\lambda_3$}%
\psfrag{s16}[l][l]{\color[rgb]{0,0,0}\small$\lambda_4$}%
\psfrag{s17}[l][l]{\color[rgb]{0,0,0}\small$\lambda_5$}%
\psfrag{s18}[l][l]{\color[rgb]{0,0,0}\small$\lambda_6$}%
\psfrag{s19}[l][l]{\color[rgb]{0,0,0}\small$\lambda_7$}%
\psfrag{s20}[l][l]{\color[rgb]{0,0,0}\small$\lambda_8$}%
\psfrag{s21}[l][l]{\color[rgb]{0,0,0}\small$\lambda_9$}%
\psfrag{s22}[l][l]{\color[rgb]{0,0,0}\small$\lambda_{10}$}%
\psfrag{s23}[l][l]{\color[rgb]{0,0,0}\small$\lambda_{11}$}%
%
% xticklabels:
\psfrag{x01}[t][t]{\footnotesize$5.0$}%
\psfrag{x02}[t][t]{\footnotesize$5.5$}%
\psfrag{x03}[t][t]{\footnotesize$6.0$}%
\psfrag{x04}[t][t]{\footnotesize$6.5$}%
\psfrag{x05}[t][t]{\footnotesize$7.0$}%
\psfrag{x06}[t][t]{\footnotesize$7.5$}%
%
% yticklabels:
\psfrag{v01}[r][r]{\footnotesize$-0.25$}%
\psfrag{v02}[r][r]{\footnotesize$-0.20$}%
\psfrag{v03}[r][r]{\footnotesize$-0.15$}%
\psfrag{v04}[r][r]{\footnotesize$-0.10$}%
\psfrag{v05}[r][r]{\footnotesize$-0.05$}%
\psfrag{v06}[r][r]{\footnotesize$0.00$}%
\psfrag{v07}[r][r]{\footnotesize$0.05$}%
\psfrag{v08}[r][r]{\footnotesize$0.10$}%
\psfrag{v09}[r][r]{\footnotesize$0.15$}%
\psfrag{v10}[r][r]{\footnotesize$0.20$}%
%
% Figure:
\includegraphics[width=0.49\linewidth]{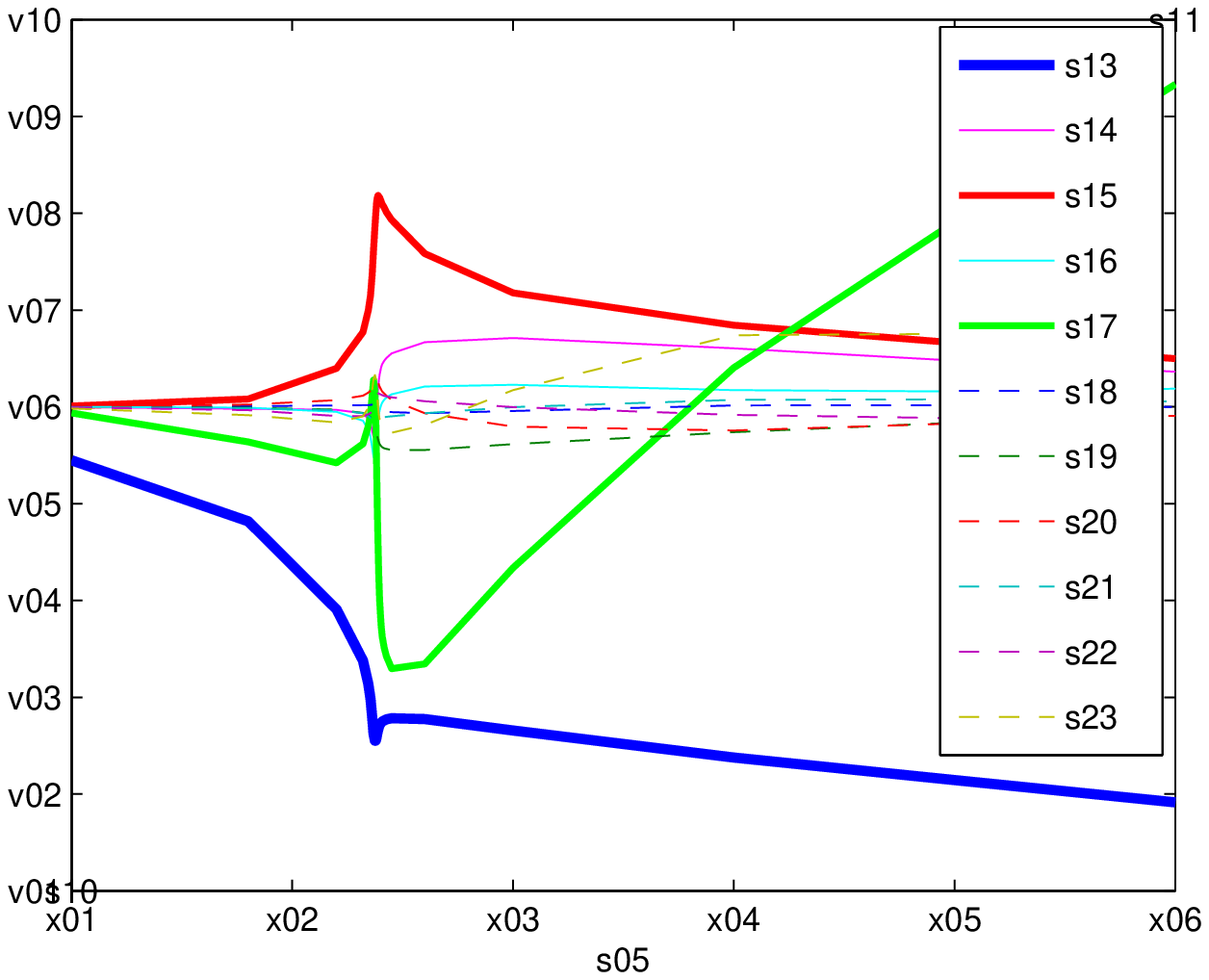}%
\end{psfrags}%
%
% End couplingsNN.tex

% This file is generated by the MATLAB m-file laprint.m. It can be included
% into LaTeX documents using the packages graphicx, color and psfrag.
% It is accompanied by a postscript file. A sample LaTeX file is:
%    \documentclass{article}\usepackage{graphicx,color,psfrag}
%    \begin{document}\input{polNextOrder}\end{document}
% See http://www.mathworks.de/matlabcentral/fileexchange/loadFile.do?objectId=4638
% for recent versions of laprint.m.
%
% created by:           LaPrint version 3.16 (13.9.2004)
% created on:           17-Apr-2007 10:51:29
% eps bounding box:     15 cm x 11.25 cm
% comment:              
%
\begin{psfrags}%
\psfragscanon%
%
% text strings:
\psfrag{s05}[t][t]{\color[rgb]{0,0,0}\setlength{\tabcolsep}{0pt}\begin{tabular}{c}\small$\beta$\end{tabular}}%
\psfrag{s06}[b][r][1][270]{\color[rgb]{0,0,0}\setlength{\tabcolsep}{0pt}\begin{tabular}{c}\small$\ev{\abs{P}}$\end{tabular}}%
\psfrag{s10}[][]{\color[rgb]{0,0,0}\setlength{\tabcolsep}{0pt}\begin{tabular}{c} \end{tabular}}%
\psfrag{s11}[][]{\color[rgb]{0,0,0}\setlength{\tabcolsep}{0pt}\begin{tabular}{c} \end{tabular}}%
\psfrag{s12}[l][l]{\color[rgb]{0,0,0}\scriptsize NNN}%
\psfrag{s13}[l][l]{\color[rgb]{0,0,0}\scriptsize Yang-Mills}%
\psfrag{s14}[l][l]{\color[rgb]{0,0,0}\scriptsize NN $\ord\!(\beta^{3\Nt})$}%
\psfrag{s15}[l][l]{\color[rgb]{0,0,0}\scriptsize NNN}%
%
% xticklabels:
\psfrag{x01}[t][t]{$0$}%
\psfrag{x02}[t][t]{$0.1$}%
\psfrag{x03}[t][t]{$0.2$}%
\psfrag{x04}[t][t]{$0.3$}%
\psfrag{x05}[t][t]{$0.4$}%
\psfrag{x06}[t][t]{$0.5$}%
\psfrag{x07}[t][t]{$0.6$}%
\psfrag{x08}[t][t]{$0.7$}%
\psfrag{x09}[t][t]{$0.8$}%
\psfrag{x10}[t][t]{$0.9$}%
\psfrag{x11}[t][t]{$1$}%
\psfrag{x12}[t][t]{\footnotesize$4.5$}%
\psfrag{x13}[t][t]{\footnotesize$5.0$}%
\psfrag{x14}[t][t]{\footnotesize$5.5$}%
\psfrag{x15}[t][t]{\footnotesize$6.0$}%
\psfrag{x16}[t][t]{\footnotesize$6.5$}%
\psfrag{x17}[t][t]{\footnotesize$7.0$}%
\psfrag{x18}[t][t]{\footnotesize$7.5$}%
\psfrag{x19}[t][t]{\footnotesize$8.0$}%
%
% yticklabels:
\psfrag{v01}[r][r]{$0$}%
\psfrag{v02}[r][r]{$0.1$}%
\psfrag{v03}[r][r]{$0.2$}%
\psfrag{v04}[r][r]{$0.3$}%
\psfrag{v05}[r][r]{$0.4$}%
\psfrag{v06}[r][r]{$0.5$}%
\psfrag{v07}[r][r]{$0.6$}%
\psfrag{v08}[r][r]{$0.7$}%
\psfrag{v09}[r][r]{$0.8$}%
\psfrag{v10}[r][r]{$0.9$}%
\psfrag{v11}[r][r]{$1$}%
\psfrag{v12}[r][r]{\footnotesize$0.0$}%
\psfrag{v13}[r][r]{\footnotesize$0.5$}%
\psfrag{v14}[r][r]{\footnotesize$1.0$}%
\psfrag{v15}[r][r]{\footnotesize$1.5$}%
\psfrag{v16}[r][r]{\footnotesize$2.0$}%
\psfrag{v17}[r][r]{\footnotesize$2.5$}%
\psfrag{v18}[r][r]{\footnotesize$3.0$}%
%
% Figure:
\includegraphics[width=0.49\linewidth]{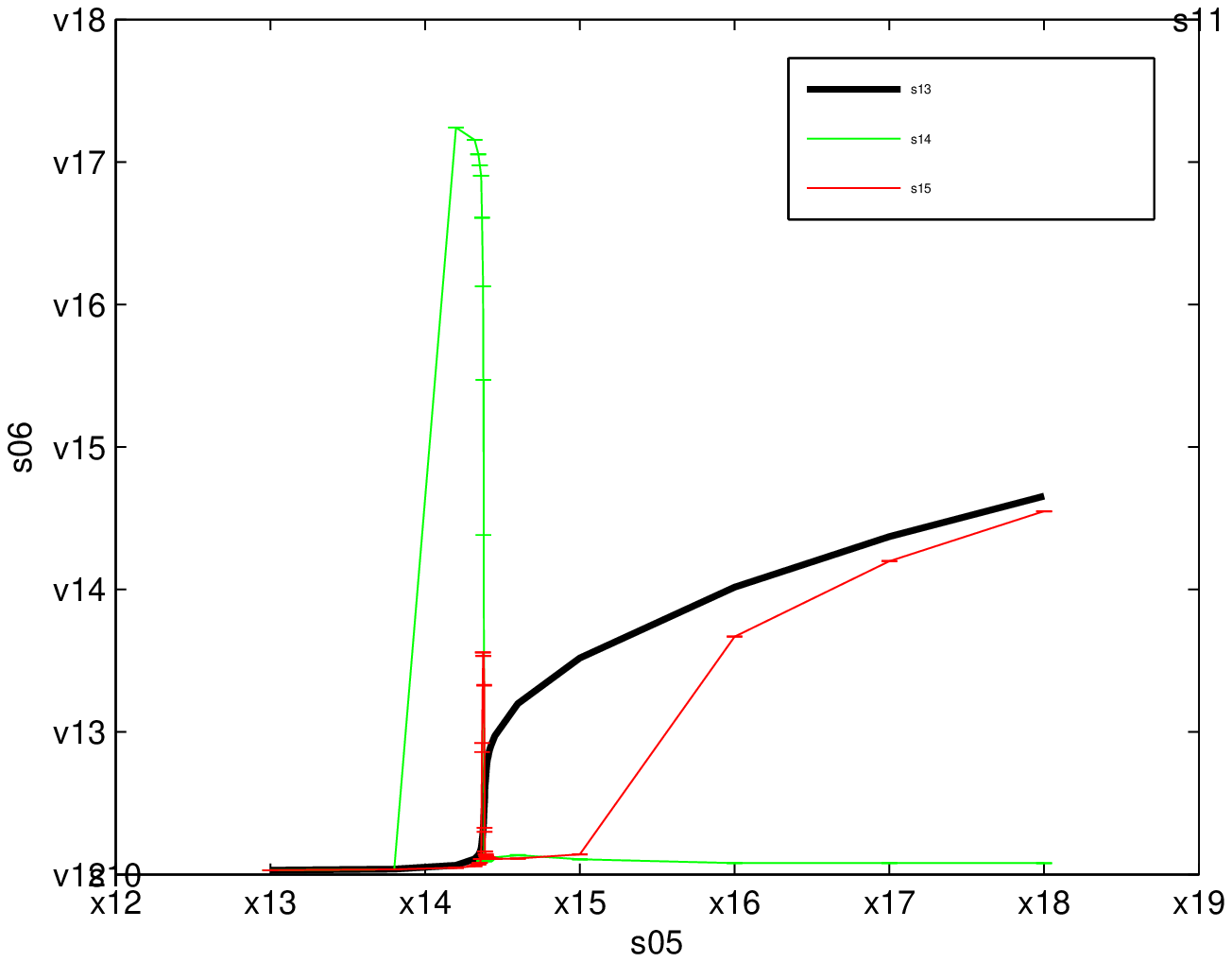}%
\end{psfrags}%
%
% End polNextOrder.tex

\caption{\label{fig:nnn} Couplings obtained for NN interactions
up to $\ord(\beta^{3\Nt})$ \textsl{(left)} and comparison of higher order
(next-to-NN) effective theories \textsl{(right)}.}
\end{figure}

Simulations with algebraic SDE on a $16^3\!\times\! 4$-lattice
allowed to determine up to $11$ effective couplings as
displayed in Fig.~\ref{fig:nnn} (left panel). The dominant
terms in the effective actions are
\begin{equation}
%\begin{aligned}
  S_1 = \sum_{\nN{\vcx\vcy}}(\chi_{10}(\trP_\vcx)\chi_{01}(\trP_\vcy)+\coco),\quad
  S_3 = \sum_{\nN{\vcx\vcy}}\chi_{11}(\trP_\vcx)\chi_{11}(\trP_\vcy), \quad
  S_5 = \sum_{\vcx}\chi_{11}(\trP_\vcx) ,
%  \end{aligned}
\end{equation}
i.e.\ two NN hopping terms and one single-site (`potential')
term.

Finally we have extended the IMC procedure to deal with NN and
next-to-NN terms up to order $\ord(\beta^{3\Nt})$ in the strong
coupling expansion. This results in an \emph{unstable behavior}
in the rendering of observables which may be traced to the
discontinuities associated with the first order phase
transition (Fig.~\ref{fig:nnn}, right panel).

% ===========================================
\section{Conclusions}
\myindent
$SU(3)$ Polyakov loop models have a surprisingly rich phase
structure when the effective couplings are allowed to vary
unrestrictedly. Upon comparing critical exponents for the
second-order antiferromagnetic phase transition we have seen
that the $SU(3)$ Polyakov loop model is in the same
universality class as the $\bbZ_3$ Potts model. The
near-perfect agreement between mean-field and Monte-Carlo
results is due to the fact that the model has a $d=3$
tricritical point. Matching the Polyakov loop models to $SU(3)$
YM theory via IMC leads to stable results only for small
lattices and a low number of couplings. Relaxing these
restrictions leads to instabilities obscuring, in particular,
the location of critical couplings. This behavior is due to the
first order nature of the phase transition in lattice
gluodynamics. Results for $SU(2)$ YM theory
\cite{Heinzl:2005xv}, on the other hand, show that IMC is
applicable for systems with second order transitions and leads
to stable results.

% ===========================================
\acknowledgments
\myindent
TK acknowledges support by the Konrad-Ade\-nauer-Stiftung
e.V. and CW by
the Studienstiftung des deutschen Volkes .
This work has been supported by the DFG grant Wi 777/8-2.

% ===========================================

\end{document}